# Smeared propagators for lattice hadron spectroscopy

G.M. de Divitiis, R. Frezzotti, M. Guagnelli, M. Masetti
and R. Petronzio

Dipartimento di Fisica, Università di Roma *Tor Vergata*

and

INFN, Sezione di Roma II

Viale della Ricerca Scientifica, 00133 Roma, Italy

February 13, 1995

**Abstract**

We propose to replace ordinary propagators in lattice operator correlations entering the determination of hadron masses with space-time smeared propagators. These are defined as the inverse of the quadratic operator in the fermion action times its hermitian conjugate. We obtain a cleaner determination of hadron masses, comparable to the effect of using smeared operators.

The determination of hadron masses in lattice QCD simulations is based on the study of the correlation of two operators at different points with appropriate quantum numbers. Due to translation invariance the correlation depends only upon the relative distance: after integrating over the space components, it depends upon the time difference and, when this becomes sufficiently large, it decreases exponentially with a slope equal to the mass of lightest hadron with the specified quantum numbers. The expression of the correlation at a generic time can be obtained by inserting a complete set of states:

$$\int d^3x \langle 0|\mathcal{O}_1(0)\mathcal{O}_2(x)|0\rangle = \sum_n \langle 0|\mathcal{O}_1(0)|n\rangle\langle n|\mathcal{O}_2(0)|0\rangle e^{-m_n t} \underset{t \text{ large}}{\propto} e^{-m_0 t} \qquad (1)$$

where $m_0$ is the mass of the lightest intermediate state. The quality of the correlation depends upon the matrix element of the chosen operator between the vacuum and this state: it is good when the contribution of this state dominates, leading to a precocious asymptotic behaviour. With *pointlike* operators it takes several lattice points before the asymptotic behaviour is established, restricting the determination of the hadron mass to a region of large times where the signal is very small.

Various improvements have been proposed in the literature [1], all based on the use of smeared sources obtained as a superposition of operators where the fundamental fields are taken in a *smearing* region with weights which can be optimized. Assembling coloured fields at different points requires a parallel transport to preserve gauge invariance or fixing the gauge. In the first case one has some degree of arbitrariness and time spent in the construction of the sum of paths for the transport, in the second case some noise is introduced by the gauge fixing procedure.

In this letter we propose a new solution which makes use of pointlike operators but modifies the propagators which appear in the correlation after integration over fermions. For example, the determination of the pion mass can be obtained by choosing the operators $\mathcal{O}_1$ and $\mathcal{O}_2$ equal to $\bar{\psi}\gamma_5\psi$ in eq. 1 which, after fermion integration, becomes:

$$\begin{aligned} \mathcal{C}(t) &= \int d^3x \langle 0|\mathrm{T}[\bar{\psi}(0)\gamma_5\psi(0)\,\bar{\psi}(x)\gamma_5\psi(x)]|0\rangle \\ &= \int \mathcal{D}A e^{-S[A]} \int d^3x \mathrm{Tr}\{G(0,x)\gamma_5 G(x,0)\gamma_5\} \end{aligned} \qquad (2)$$

where $G$ are the fermion propagator in a fixed gauge field configuration $\{A\}$ and only the flavour non singlet correlation is retained.
The idea is to replace the usual propagator with one which leads to the same hadron spectrum but which is more suppressed with increasing quark off-shellness. Excited states are expected to be correspondingly more suppressed. The standard propagator is the inverse of the quadratic operator appearing in the Wilson action:

$$M(x,y) = \delta_{x,y} - K\sum_{\mu=0}^{3}\left[U_\mu(x)(1-\gamma_\mu)\delta_{x+\mu,y} + U_\mu^\dagger(x-\mu)(1+\gamma_\mu)\delta_{x-\mu,y}\right]. \qquad (3)$$

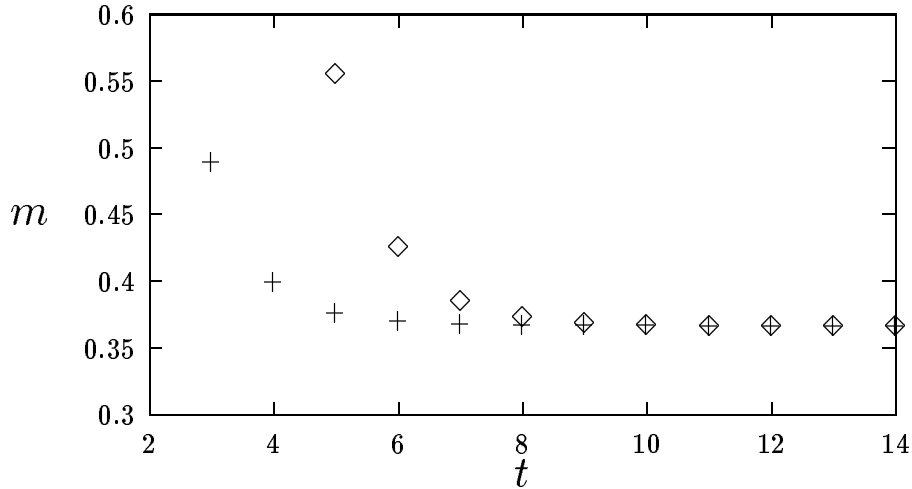

Figure 1: The effective mass for the pion correlation in the free quarks case. The data + correspond to the smeared case.

The new propagator is taken as the inverse of the operator $M^\dagger M$ and can be seen as a *smearing*, also in time, of the original propagator.
For free quarks in the continuum this just corresponds to replacing the usual euclidean fermion propagator

$$\frac{m - i\not{p}}{p^2 + m^2} \qquad \text{with} \qquad \frac{1}{p^2 + m^2}. \tag{4}$$

In this case it is easy to obtain from eq. 2 the analytic form of the correlation which is expected for large times to be dominated by the lowest energy state of two free quarks of mass $m$:

$$\mathcal{C}(t) \propto \int_m^\infty dE \ E^\alpha \sqrt{E^2 - m^2} e^{-2Et} \tag{5}$$

The value of the parameter $\alpha$ is 1 in the standard case and $-1$ with the smeared propagator, leading in the latter case to a stronger suppression in the momentum integral over off-shell intermediate states. In figure 1 it is shown the effective mass on the lattice for the two cases for free Wilson fermions.

For the interacting case, before proceeding to a comparison of the correlations, we want to show that the ones obtained from smeared propagators are indeed dominated for large times by the lowest lying state.

The proof uses the following relation between the standard and the smeared propagator:
$$(M^\dagger M)^{-1} = M^{-1}\gamma_5 M^{-1}\gamma_5 \qquad (6)$$
The expression for the correlation with meson sources and smeared propagators can be rewritten as:
$$\int d^3x \int d^4y d^4z \; M^{-1}{}_{xy}\gamma_5 M^{-1}{}_{y0}\gamma_5 \Gamma M^{-1}{}_{0z}\gamma_5 M^{-1}{}_{zx}\gamma_5 \Gamma \qquad (7)$$
where for pion $\Gamma = \gamma_5$, and reinterpreted as the following new correlation:
$$\int d^3x \int d^4y d^4z \langle 0|T\left[\pi_{12}(y)\Gamma_{23}(0)\pi_{34}(z)\Gamma_{41}(x)\right]|0\rangle$$
$$\begin{cases} \pi_{ij} = \bar{\psi}_i \gamma_5 \psi_j \\ \Gamma_{ij} = \bar{\psi}_i \gamma_5 \Gamma \psi_j \end{cases} \qquad (8)$$

where the quark fields entering the correlation are labelled with a new flavour index; each new flavour has the same mass, standard propagators, but does not give rise to additional fermion loops and it must be considered quenched. The distinction between flavours is needed in the Wick expansion to produce only the contractions represented in eq. 7. The spectrum contains new states exactly degenerate with the standard ones. By summing over all different time orderings, using translation invariance and inserting a complete set of states one obtains the final expression for the continuum case in a infinite volume:

$$\int d^3x \int d^4y d^4z \langle 0|T\left[\pi_{12}(y)\Gamma_{23}(0)\pi_{34}(z)\Gamma_{41}(x)\right]|0\rangle =$$
$$= \sum_{a,b,c} \left\{ \frac{e^{-m_c t_x} - e^{-m_a t_x}}{(m_c - m_a)(m_c - m_b)} - \frac{e^{-m_b t_x} - e^{-m_a t_x}}{(m_c - m_b)(m_b - m_a)} \right\} \times$$
$$\times \langle 0|\Gamma_{41}(0)|a\rangle\langle a|\pi_{34}(0)|b\rangle\langle b|\pi_{12}(0)|c\rangle\langle c|\Gamma_{23}(0)|0\rangle$$
$$+ \frac{e^{-m_b t_x} - e^{-m_c t_x}}{m_a(m_c - m_b)} \langle 0|\pi_{34}(0)|a\rangle\langle a|\Gamma_{41}(0)|b\rangle\langle b|\pi_{12}(0)|c\rangle\langle c|\Gamma_{23}(0)|0\rangle$$
$$+ \frac{e^{-m_a t_x} - e^{-m_b t_x}}{m_c(m_b - m_a)} \langle 0|\Gamma_{41}(0)|a\rangle\langle a|\pi_{34}(0)|b\rangle\langle b|\Gamma_{23}(0)|c\rangle\langle c|\pi_{12}(0)|0\rangle$$
$$+ \frac{e^{-m_c t_x}}{m_a m_b} \langle 0|\pi_{34}(0)|a\rangle\langle a|\pi_{12}(0)|b\rangle\langle b|\Gamma_{41}(0)|c\rangle\langle c|\Gamma_{23}(0)|0\rangle$$
$$+ \frac{e^{-m_a t_x}}{m_c m_b} \langle 0|\Gamma_{41}(0)|a\rangle\langle a|\Gamma_{23}(0)|b\rangle\langle b|\pi_{34}(0)|c\rangle\langle c|\pi_{12}(0)|0\rangle$$
$$+ \frac{e^{-m_b t_x}}{m_a m_c} \langle 0|\pi_{34}(0)|a\rangle\langle a|\Gamma_{41}(0)|b\rangle\langle b|\Gamma_{23}(0)|c\rangle\langle c|\pi_{12}(0)|0\rangle$$
$$+ \{\pi_{34} \leftrightarrow \pi_{12}\} \qquad (9)$$

The main difference with respect to the usual correlations is that the parity of the intermediate states is not determined by the one of the operator chosen for the correlation: both parity states contribute because of the insertion of the parity commuter operator $\bar{\psi}\gamma_5\psi$. For large times the lowest lying state dominates: however, in the case where the source operators are of even parity, the coefficient of the exponential of the pseudoscalar state is the sum of a constant term and of one increasing linearly with time and leading to a slow drift of the effective mass calculated assuming a constant coefficient, which reaches the true value only asymptotically. This is not the case for parity odd operators, where such a behaviour affects the exponential of the heavier scalar state. In general, such a pathology occurs when there are two degenerate intermediate states in the particular time ordering where the parity commuting operators occurr at times between those of the two source operators.

In figures 2 and 3 we compare the $\pi$ and $\rho$ effective masses [1] obtained with standard and smeared propagators on a $16^3 \times 32$ lattice at $\beta = 5.7$ for quenched Wilson fermions at $K = 0.161$: in the latter case the *plateau* appears at smaller values of the time distance. The points referring to scalar and pseudovector sources are obtained from a naive analysis ignoring the pathology of the term linear with time. A proper fit which allows for this behaviour leads to values of the mass compatible with those of the standard and smeared propagators with pseudoscalar and vector sources.

For baryon states, the technique which allows to show that the correlation is indeed dominated by the lowest lying state of any parity is similar to the meson case. The correlation of the two proton operators of the form:

$$\int d^3x \langle 0|T[\Psi^\delta(x)\bar{\Psi}^{\delta'}(0)]|0\rangle$$

$$\text{with } \Psi^\delta(x) = [u_a^T(x)(C\gamma_5)d_b(x)] \, u_c^\delta(x) \, \epsilon^{abc},$$

$$\bar{\Psi} = \Psi^\dagger \gamma_0 \text{ and } C \text{ the charge conjugation matrix} \quad (10)$$

can be transformed into:

$$\int d^3x \int d^4y d^4z d^4w \langle 0|T\left[\Psi_1^\delta(x) \, \pi_{12}^d(y)\pi_{12}^u(z)\pi_{12}^u(w) \, (\bar{\Psi}_2(0)\gamma_5)^{\delta'}\right]|0\rangle$$

$$\text{where} \quad \Psi_i^\delta(x) = [u_{a,i}^T(x)(C\gamma_5)d_{b,i}(x)] \, u_{c,i}^\delta(x) \, \epsilon^{abc}$$

$$\text{and} \quad \pi_{ij}^f = \bar{f}_i \gamma_5 f_j \quad (11)$$

using the relation in eq. 6 between the space-time smeared propagators and the standard ones. By inserting a complete set of states and performing the integrals over times as in the meson correlation case one gets a sum of contributions from baryons of both parities. An interesting property of such a correlation is the symmetry with respect to the center of the lattice, which allows to further average statistically the signal. This can be shown by using the representation of the correlation in terms of propagators with their periodicity properties and by considering the contributions coming from the correlations forward and backward in time.

---

[1] The effective mass is obtained from an analysis of the correlation which assumes only one intermediate state at any time.

In the case of baryons, it is not possible to eliminate by a suitable choice of the source the contribution where there are two degenerate intermediate states with the lowest mass because of the insertion of *three* parity commuting operators. The determination of the effective mass is then biased, at moderate times, by the presence of a coefficient linear in time, although in this case one gets a smaller percentage correction with respect to the pion case.
A solution to this problem is the choice of a single smeared propagator and two standard ones. The symmetry of the correlation is lost in this case, but it can be shown that, for the best case where non relativistic sources are used, the signal from the correlation backward in time is very small and leaves a sizeable "fiducial" region for the determination of the effective mass *plateau*.

The final results for baryons are shown in figure 4. In the first plot we show the effective mass for the following non relativistic proton operator:

$$\Psi_\pm^\delta(x) = [u_a^T(x)(P_\pm C\gamma_5)d_b(x)]\, u_c^\delta(x)\, \epsilon^{abc} \qquad \text{with} \quad P_\pm = \frac{1}{2}(1 \pm \gamma_0) \qquad (12)$$

with standard propagators on the up quarks and a smeared one on the down quark compared with the cases with three standard or three smeared propagators. In figure 5 we show the absolute value of the real part of the correlations for the same three cases. The points where the correlation changes sign are affected by large statistical fluctuations and are not shown in the figure.

The use of space-time smeared propagators has been shown to improve significantly the quality of the correlations from which hadron masses are determined. The extra computing effort is rather limited and needed anyway if the conjugate gradient algorithm is used in the inversion: standard propagator correlations are obtained by smearing the quark fields in the operator creating the initial state with the local operator $M^\dagger$. The technique presented in this letter can be further improved with the standard operator smearing.

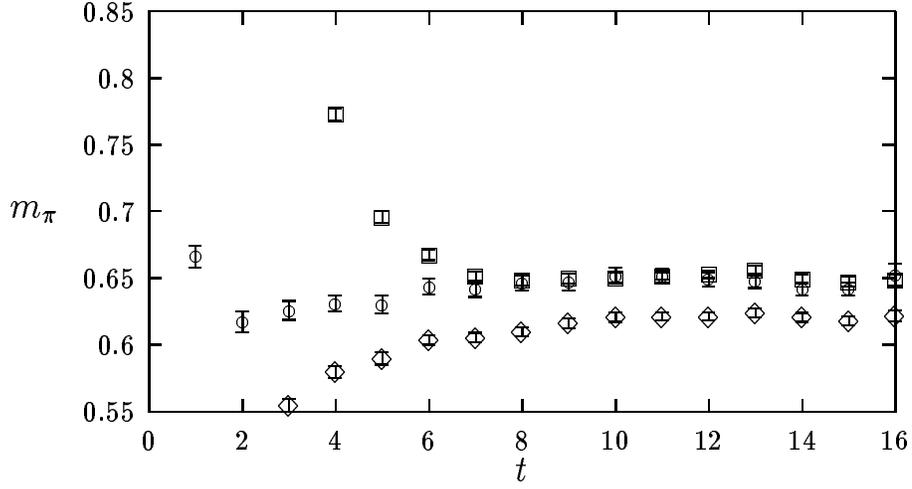

Figure 2: The pion effective mass on a $16^3 \times 32$ lattice at $\beta = 5.7$ for three different cases: ($\square$) standard propagators, ($\circ$) smeared propagators with a pseudoscalar source operator, ($\diamond$) smeared propagators with a scalar source operator.

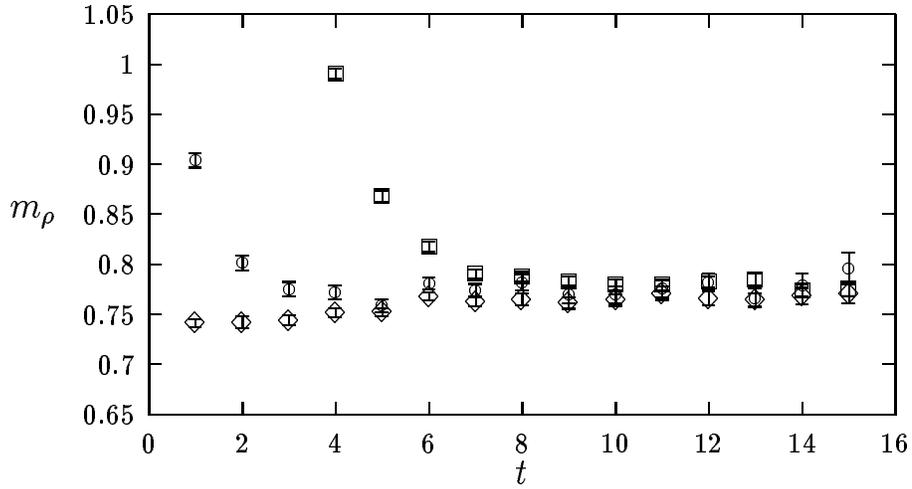

Figure 3: The $\rho$ effective mass for three different cases: ($\square$) standard propagators, ($\circ$) smeared propagators with a vector source operator, ($\diamond$) smeared propagators with a pseudovector source operator.

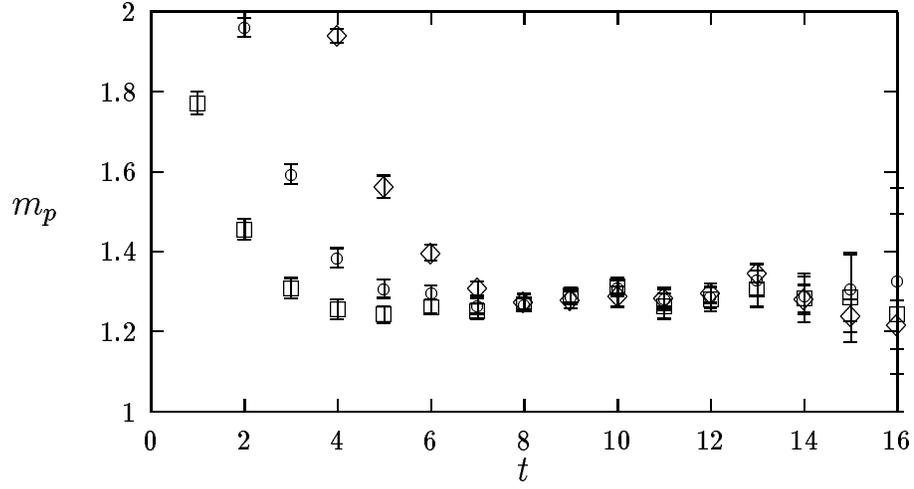

Figure 4: The proton effective mass for the cases: ($\diamond$) three standard propagators, ($\circ$) one smeared and two standard propagators, ($\square$) three smeared propagators.

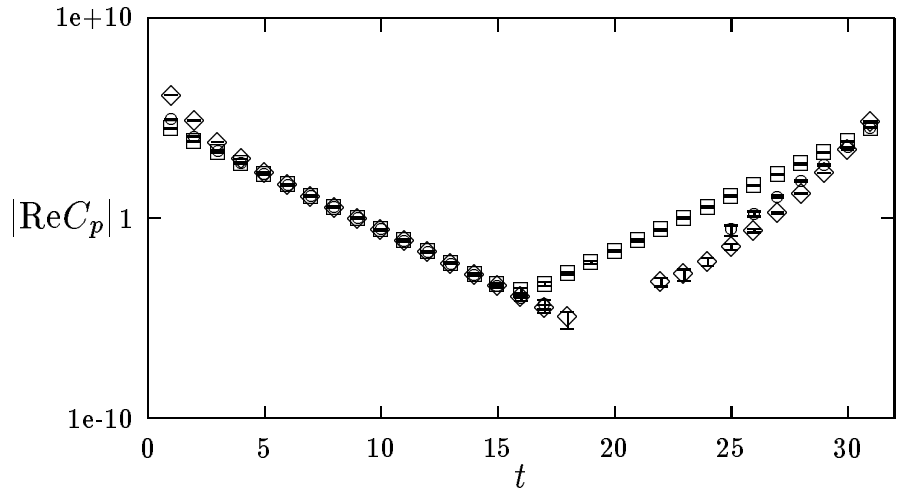

Figure 5: The absolute value of the real part of the correlation for the three cases defined in the previous figure. Points where the signal changes sign and is affected by large statistical error are omitted.

8